\documentclass{article}%
\usepackage{amsfonts}
\usepackage{amsmath}
\usepackage{amssymb}
\usepackage{graphicx}%
\begin{document}

\title{Addendum to: Area density of localization-entropy I }
\author{Bert Schroer\\CBPF, Rua Dr. Xavier Sigaud 150 \\22290-180 Rio de Janeiro, Brazil\\and Institut fuer Theoretische Physik der FU Berlin, Germany}
\date{January 2007}
\maketitle

\begin{abstract}
This addendum adds some comments to a published paper on area density of
localization-entropy. The setting of holography and localization-entropy is
generalized to double cones and black holes with bifurcate Killing horizons.


\end{abstract}

\section{A brief resum\'{e} of lightfront holography}

The algebraic setting of QFT the \textit{holographic projection}\footnote{The
terminology goes back to t'Hooft who coined it for a property which he
expected to be an important attribute of a future theory of quantum gravity
(QG). In the present context it refers to a property of localized quantum
matter which is not a new requirement but rather follows from the standard
causal locality and spectral properties of QFT. Although it has no direct
relation with the still elusive QG, it explains some of the properties (e.g.
the area proportionality of vacuum fluctuation caused entropy) which have been
attributed to QG.} on null-surfaces replaces the system of subalgebras of a
given localized algebra (associated with bulk subregions) by a new system of
subalgebras indexed with subregions of the causal horizon. In \cite{S1} this
construction was presented for the holography onto the lightfront
null-surface. It is precisely the relation of \textit{localization} with
\textit{vacuum-fluctuation-caused thermal properties} and their mathematical
connection with \textit{modular operator algebra theory} which forces one to
substitute the old incomplete (and not entirely correct) ideas around
\textquotedblleft lightcone quantization\textquotedblleft\ with
\textit{lightfront holography} in the sense of this paper.

The starting point is the previously presented fact that a lightfront $LF$ is
the linear extension of the upper horizon $\partial W$ of a fixed wedge $W$
region (all regions are open) which reveals the thermal manifestations of
vacuum polarization and permits to use them in a constructive way. The
\textit{causal shadow property} of the algebraic approach (stating that an
operator algebra associated with a region $\mathcal{O}$ is equal to that
associated with the generally larger causally completed region $\mathcal{O}%
^{\prime}{}^{\prime}$) identifies the operator algebra belonging to the
characteristic surface $\partial W$ with that of its causally associated bulk
region $W$%
\begin{equation}
\mathcal{A}(\partial W)\equiv\mathcal{A}(W)\label{hol}%
\end{equation}
But it does not provide algebraic localization properties for subregions of
these semi-global operator algebras. The aim of hologaphy is to construct a
coherent system of subalgebras $\mathcal{A(O)}\subset\mathcal{A}(\partial W)$
(where from now the letter $\mathcal{O}$ denotes finitely extended regions on
$LF)$ and this local refinement of the lightfront algebra $\mathcal{A}(LF)$ is
the source of the computational power of lightfront holography versus the old
\textquotedblleft lightcone quantization\textquotedblleft$.$ In the spirit of
Leibniz's identification of spacetime as an \textit{ordering device for
matter}, holography is a rather radical change of that device and of the
physical interpretation, but it maintains the Hilbert space of the bulk theory
as well as those subalgebras in the bulk which are causal shadows of regions
on $LF;$ these shadow regions consist of those wedges $W$ whose upper horizons
$\partial W$ are halfplanes of $LF.$

The often raised question whether thermal manifestations of localization, in
particular localization-entropy, are phenomena associated with $W$ or
$\partial W$ is somewhat academic in view of the identity (\ref{hol}). It
would be reasonable to maintain that from the viewpoint of the material
substrate this distinction is not intrinsic but rather pendent on one's choice
of the spacetime ordering device. Bulk regions (black holes) and their causal
(event) horizons are mainly distinguished in this sense of Leibniz. Holography
is a more radical form of the recent statement in the setting of QFT in CST
\cite{B-F-V} that local quantum physics in isometric regions of different
universes is isomorphic. All these statements are special cases of the still
somewhat mysterious possibility of encoding all material properties (including
the distiction between different kinds of quantum matter) as well as the
geometric spacetime positioning into appropriately defined modular inclusions
and intersection of a finite number of copies of the same abstract operator
algebra (a \textit{monade,} or in the terminology of operator algebras a
hyperfinite type III$_{1}$ von Neumann factor algebra). This may be viewed as
an abstract ordering device in Hilbert space which generates the richness of
the \textit{material world and spacetime from the positioning of a finite
number of monades.}

The symmetry group $\mathcal{G}$ of $LF$ is known to be a 7-parametrig
subgroup of the 10-parametric Poincar\'{e} group which is generated by of
lightray translations, dilations (the $LF$ projections of the W-preserving
boost) and certain linear transformations of the coordinate directions in the
transverse space $\mathbb{R}^{2}\subset LF$ (resulting from the $LF$
projection of the 3-parameter Wigner little group which leaves the lightray
invariant) together with transverse translations. The local subalgebras
$\mathcal{A(O)}$ with $\mathcal{O}\subset LF$ of the holographic projection
are obtained by successive steps involving intersections of wedge algebras
obtained from $W$ by the application of $\mathcal{G}.~$By iterating this
process of taking intersections and acting with $\mathcal{G},$ one obtains a a
coherent net of algebras $\mathcal{A(O)}$ indexed by arbitrarily small regions
$\mathcal{O\subset~}LF$.

The results in \cite{S1} may be summarized as follows

\begin{enumerate}
\item The system of $LF$ subalgebras $\left\{  \mathcal{A(O)}\right\}
_{\mathcal{O\subset}LF}$ tensor-factorizes transversely with the vacuum being
free of transverse entanglement
\begin{align}
&  \mathcal{A(O}_{1}\mathcal{\cup O}_{2}\mathcal{)}=\mathcal{A(O}%
_{1}\mathcal{)\otimes A(O}_{2}\mathcal{)},\text{ }\mathcal{(O}_{1}%
\mathcal{)}_{\perp}\cap\mathcal{(O}_{2}\mathcal{)}_{\perp}=\emptyset
\label{fac}\\
&  \left\langle \Omega\left\vert \mathcal{A(O}_{1}\mathcal{)\otimes A(O}%
_{2}\mathcal{)}\right\vert \Omega\right\rangle =\left\langle \Omega\left\vert
\mathcal{A(O}_{1}\mathcal{)}\left\vert \Omega\right\rangle \left\langle
\Omega\right\vert \mathcal{A(O}_{2}\mathcal{)}\right\vert \Omega\right\rangle
\nonumber
\end{align}

\item Extensive properties as entropy and energy on $LF$ are proportional to
the extension of the transverse area.

\item The area density of localization-entropy in the vacuum state for a
system with sharp localization on $LF$ diverges logarithmically
\begin{equation}
s_{loc}=\lim_{\varepsilon\rightarrow0}\frac{c}{6}\left\vert ln\varepsilon
\right\vert +...\label{ent}%
\end{equation}
where $\varepsilon$ is the size of the interval of \textquotedblleft
fuzziness\textquotedblright\ of the boundary in the lightray direction which
one has to allow in order for the vacuum polarization cloud to attenuate and
the proportionality constant $c$ is (at least in examples) the central
extension parameter of the Witt-Virasoro algebra.
\end{enumerate}

The following comments about these results are helpful in order to appreciate
some of the physical consequences as well as possibilities of extension to
more general null-surfaces.

The transverse factorization with respect to the vacuum state is the
consequence of a general structural theorem \cite{Bo} of Local Quantum Physics
(LQP). The latter states that two operator algebras $\mathcal{A}_{i}\subset
B(H),$ $i=1,2$ with $\left[  A_{1},U(a)\mathcal{A}_{2}U(a)^{\ast}\right]  =0$
$\forall a,$ and $U(a)$ a translation with nonnegative generator which
fulfills the cluster factorization property (i.e. asymptotic factorization in
correlation functions for infinitely large cluster separations) with respect
to a unique $U(a)$-invariant state vector $\Omega$ automatically satisfy the
stronger tensor factorization property (a strong form of statistical
independence) in the sense of the above statement \textbf{1 }which\textbf{ }in
turn implies\textbf{ }clustering and commutativity. In the case at hand the
tensor factorization follows as soon as the regions have no transverse
overlap\footnote{Locality in both directions shows that the lightlike
translates $\left\langle \Omega\left\vert AU(a)B\right\vert \Omega
\right\rangle $ are boundary values of entire functions and the cluster
property together with Liouville's theorem gives the factorization.}. 

Evidently this theorem has far-reaching consequences for algebraic nets
indexed by subregions on null-surfaces in curved spacetime. Besides the
quantum mechanical nature in transverse directions on $LF$ it guaranties the
absence of corresponding vacuum polarizations in transverse directions on the
mantle of a lightcone or of a compact double cone in conformal QFTs for which
these regions are obtained from wedges by conformal transformations. The
asymptotic cluster factorization holds for spacelike separations and in a
weaker form also for lightlike cluster separations in the ambient Minkowski
spacetime bulk \cite{Com}.

Let $W$ be the $x_{0}-x_{3}$ wedge in Minkowski spacetime which is left
invariant by the $x_{0}-x_{3}$ Lorentz-boosts. Consider a family of wedges
$W_{a}$ which are obtained by sliding the $W$ along the $x_{+}=x_{0}+x_{3}$
lightray by a lightlike translation $a>0$ into itself. The set of spacetime
points on $LF$ consisting of those points on $\partial W_{a}$ which are
spacelike to the interior of $W_{b}$ for $b>a$ is denoted by $\partial
W_{a,b};$ it contains all points $x_{+}\in(a,b)$ with an unlimited transverse
part $x_{\perp}\in R^{2}$. These regions are two-sided transverse "slabs" on
$LF$. To get to intersections of finite size one may \textquotedblleft
tilt\textquotedblright\ these slabs by the action of certain subgroups in
$\mathcal{G}$ which change the transverse directions. Using the 2-parametric
subgroup $\mathcal{G}_{2}$ of $\mathcal{G},$ which is the restriction to $LF$
of the two \textquotedblleft translations\textquotedblright\ in the Wigner
little group (i.e. the subgroup fixing the lightray in $LF$), it is easy to
see that this is achieved by forming intersections with $G_{2}$- transformed
slabs $\partial W_{a,b}$
\begin{equation}
\partial W_{a,b}\cap g(\partial W_{a,b}),\text{ }g\in\mathcal{G}%
_{2}\label{int}%
\end{equation}
Continuing with forming intersections and unions, one can get to finite convex
regions $\mathcal{O}$ of a quite general shape.

An alternative method for obtaining holographically projected compactly
localized subalgebras $\mathcal{A(O)},\mathcal{O}\subset LF$ which does not
make use of transverse symmetries, consists in intersecting $\mathcal{A}%
(\partial W_{a})$ with suitable algebras in the bulk which are localized in a
tubular neighborhood of $\mathcal{O}$ \cite{G-L-R-V}. This is useful for
null-surfaces in curved spacetime (next section).

The nontrivial question is now whether this geometric game can be backed up by
the construction of a nontrivial net of operator algebras which are indexed by
those regions. Since a subregion on $\partial W,$ which either does not extend
to infinity in the $x_{+}$ lightray direction or lacks the two-sided
transverse extension does not cast any causal shadow\footnote{In the classical
setting this means that such characteristic data in contrast to data on
$\partial W$ ($W$ arbitrary) on LF do not define a hyperbolic propagation
problem in the ambient spacetime.}, one cannot base the nontriviality of
algebras $A(\partial W_{a,b})$ on the causal shadow property. If this algebra
would be trivial (i.e. consist of multiples of the identity), the motivation
for the use of holographic projections (which consists in obtaining a simpler
description of certain properties) would fall flat and with it the dream of
simplifying certain physical aspects via lightlike holography.

It has been customary in the algebraic approach to add structural properties
concerning intersections to the \textquotedblleft axiomatic\textquotedblright%
\ list of algebraic requirements if they can be derived in the absence of
interactions and at least formulated in the presence of interaction-caused
vacuum polarization clouds, remembering that the intrinsic characterizations
of interactions is the inexorable presence of such clouds in local states
(states by acting with a local operator onto the vacuum). This is a
generalization to the setting of algebraic QFT of a theorem known since the
early 60s that a pointlike field which is not free upon application to the
vacuum generates necessarily a vacuum polarization cloud in addition to to a
one-particle component. For noncompact localization regions as wedges there
are vacuum \textbf{p}olarization-\textbf{f}ree \textbf{g}enerators even in the
presence of interactions (the PFGs in \cite{B-B-S}). There is presently no
indication that the nontriviality of the algebraic intersections of bulk
algebras needed for the construction of the local algebras $\mathcal{A(O})$ on
the lightfront is in any way impeded by interactions. But at least there
exists a proof \cite{Lech} that within the family of two-dimensional
factorizing models the double cone intersections of two wedge algebras in the
bulk act cyclically on the vacuum. After 80 years of QFT without having been
able to secure the existence of any nontrivial model (except
superrenormalizable models with a finite wave-function renormalization), the
setting of operator algebras, in particular modular theory, has \ opened new
avenues beyond perturbation theory. The proof that the nontriviality continues
to be valid for the intersections needed in the holographic projection is only
a matter of time.    

In the following the nontriviality of holography will be shown in the absence
of interactions. It is well-known that the system of interaction-free
localized operator algebras $\mathcal{A(O)},\mathcal{O}\subset LF$ (which are
constructed with the help of Weyl algebra generators from free fields) do
indeed pass this nontriviality test for sufficiently many finite regions
$\mathcal{O}$ of interests \cite{Dries}. In fact these algebras on $LF$ have
also pointlike generators; the generating fields are the well-known
\textquotedblleft lightcone quantization\textquotedblright\ (or
\textquotedblleft$p\rightarrow\infty$ frame\textquotedblright\ method) fields
$A_{LF}$ (using the abbreviation $x_{\pm}=x^{0}\pm x^{3},~p_{\pm}=p^{0}%
+p^{3}\simeq e^{\mp\theta},~\theta$ the rapidity):
\begin{align}
&  A_{LF}(x_{+},x_{\perp})\simeq\int\left(  e^{i(p_{-}(\theta)x_{+}+ip_{\perp
}x_{\perp}}a^{\ast}(\theta,p_{\perp})d\theta dp_{\perp}+h.c.\right)
\label{LF}\\
&  \left\langle \partial_{x_{+}}A_{LF}(x_{+},x_{\perp})\partial_{x\prime_{+}%
}A_{LF}(x_{+}^{\prime},x_{\perp}^{\prime})\right\rangle \simeq\frac{1}{\left(
x_{+}-x_{+}^{\prime}+i\varepsilon\right)  ^{2}}\cdot\delta(x_{\perp}-x_{\perp
}^{\prime})\nonumber\\
&  \left[  \partial_{x_{+}}A_{LF}(x_{+},x_{\perp}),\partial_{x\prime_{+}%
}A_{LF}(x_{+}^{\prime},x_{\perp}^{\prime})\right]  \simeq\delta^{\prime}%
(x_{+}-x_{+}^{\prime})\delta(x_{+}-x_{\perp}^{\prime})\nonumber
\end{align}
where, we for the sake of brevity and structural clarity, we left out all
unimportant constant and took the lightlike derivative in order to avoid
technicalities about logarithmic zero mass correlations (a pseudo problem,
resolved in terms of test-functions as well-known from the massless field in
2-dim. QFT\footnote{It can be shown that working with the $x_{+}$ derivates
does not change the affiliated (weakly closed) operator algebras.}).
\textit{The only physical purpose of this auxiliary field is to generate a
system of local operator algebras on }$LF$; their correlation function have no
other direct relation to the physical correlation in the bulk\footnote{That
holographic pointlike generators are still related to the free fields in the
bulk in a way which resembles a \textquotedblleft
restriction\textquotedblright\ is a speciality of the linear lightfront and
does not extend to other null-surfaces. At least this aspect (\ref{LF}) of the
old "lightcone quantization" was treated correctly.}; in particular in case of
free fields the $A_{LF}$ correlation functions on $LF$ and the bulk
correlation in terms of $A$ at the same separation are very different; e.g.
the x-space bulk two-point function diverges for lightlike separation whereas
that of $A_{LF}$ stays finite and fulfills the kind of lightlike locality
known from 2-dim. conformal QFTs. 

The reverse reconstruction (the inverse holography) of the physical fields
from the $LF$ fields goes again through algebraic steps; in that case one uses
the generating property of the pointlike $A_{LF}$ which together with
(\ref{hol}) gives $\mathcal{A}(W)$ and reconstructs the system of wedge
algebras for wedges in all possible positions by applying the Poincar\'{e}
group. Finally from the net of wedge-localized algebras one obtains the
\textit{local algebraic structure of the bulk again via intersections. }There
is no direct way to obtain\textit{ } to pass from the $A_{LF}$ generating
fields to the bulk generating fields $A$ rather the latter emerge as the local
generators of these bulk (double cone) algebras for arbitrarily small
localization regions. Clearly the possibility of directly constructing
pointlike $LF$ generators hinges on the mass-shell property of free fields; in
the general interacting case the algebraic construction via intersections
seems to be unavoidable and the problem of pointlike generators remains open,
although the following generalization of the commutation relation to those of
a \textit{transversely extended chiral theory} has a high degree of
plausibility
\begin{equation}
\left[  B_{LF}^{(i)}(x_{+},x_{\perp}),B_{LF}^{(k)}(x_{+}^{\prime},x_{\perp
}^{\prime})\right]  \simeq\left\{  \sum_{l}\delta^{(n_{l}}(x_{+}-x_{+}%
^{\prime})B_{LF}^{(l)}(x_{+},x_{\perp})\right\}  \delta(x_{\perp}-x_{\perp
}^{\prime})\label{c.r.}%
\end{equation}
where in the case of transversely extended rational theories the algebraic
structure of the theory permits a characterization in terms of a finite number
of $LF$ generating fields $B_{LF}^{(i)}.$ For Wick monomials of free
fields\ without derivatives this form of the relation can also be obtained
directly by using their mass-shell representation for the individual fields in
the Wick product. But In the interacting case one should expect that the
algebraically determined net $\mathcal{A}(\mathcal{O})$ does not reproduce the
original $\partial W$ global algebra i.e. $\overline{\cup_{\mathcal{O}%
\subset\partial W}\mathcal{A}(\mathcal{O})}\varsubsetneq$ $\mathcal{A}%
(\partial W)=\mathcal{A}(W).~$In field theoretic terms the bulk fields with a
non-integer dimension are mapped into zero by algebraic holography since the
bulk spin and the bulk short distance scale dimension have to match in order
to pass the holographic lightfront projection. In this case there is the hope
to recover $\mathcal{A}(\partial W)$ from its local net components by
extending the net with the help of its superselection structure. There is also
an indication that by using nonlocal formal representations of bulk Heisenberg
fields in terms of incoming fields, one can define a more general holographic
procedure which allows to incorporate bulk fields with anomalous short
distance dimensions. The issue of holography applied to bulk fields with
fractional short distance dimensions remains an important project for the
future with the simplest case being that of two-dimensional factorizing models.

In the vast literature on \textquotedblleft light-cone
quantization\textquotedblright\ this problem as well as the question about the
connection between the bulk- and the $LF-$ locality has not been properly
addressed, and most of the computations about light-cone quantization use ad
hoc prescriptions when interactions are present. \ 

The appearance of the conformal invariant two-point functions in lightray
direction (\ref{LF}) begs the question whether the two symmetry operations
which the $LF$ system of local algebras inherits from the bulk symmetries,
namely the lightray translation and dilation (the restriction of the
$W$-boost), can always be completed to the 3-parameter Moebius group. Using
properties of the system of operator algebras $\left\{  \mathcal{A(O)}%
\right\}  _{\mathcal{O}\subset LF}$ one shows that a physically reasonable
algebraic property\footnote{This means one which can be checked for algebras
of free fields and in whose formulation interaction plays no direct role.},
namely the non-triviality and cyclicity with respect to the vacuum state (the
Reeh-Schlieder property of QFT\footnote{In the mathematical setting of
operator algebras the cyclicity of an algebra and its commutant with respect
to a vector state $\Omega$ is called the \textit{standardness} property.}
\cite{Haag}) of the algebra $\mathcal{A}(\partial W_{a,b}),$ guaranties the
existence of a third one-parametric group of unitary operator $U(Rot)$ in the
Hilbert space of the theory which leaves the vacuum invariant and (together
with the lightray dilation and translation) generates the Moebius group
\cite{G-L-W}. 

At this point the operator-algebraic setting unfolds its \textquotedblleft
magic\textquotedblright\ in that new symmetries which were not present in the
bulk theory arise in the holographic image from coherence properties of
operator algebras in certain relative positions (modular inclusions) within a
common Hilbert space. This symmetry-increase is typical for holographic
projections onto null-surfaces and makes holographic projection a simplifrying
and therefore useful tool. The above interaction-free case, which leads to an
$A_{LF}$ field which is similar to the 2-dim. abelian current algebra, admits
a much larger unitarily implemented symmetry group, namely the diffeomorphism
group of the circle. However the unitary implementers (beyond the Moebius
group) do not leave the vacuum invariant and hence are not Wigner symmetries.
These Diff(S$^{1}$) may well appear in the holographic projection of
interacting theories, even though the standard argument via a chiral
energy-momentum tensor is not available for holography onto null-surfaces. As
the vacuum-preserving chiral rotation results from algebraic properties of the
holographic projection, it is conceivable that one finds algebraic properties
which lead to the higher diffeomorphism symmetries. This would be very much in
the spirit of the new setting of local covariance \cite{B-F-V}.

It is helpful to add one more remark with respect to the first statement in
the above list. It is well-known that for spacelike separation in the bulk,
even with a finite spacelike separation of two localization regions, there is
no factorization on the vacuum vector $\Omega;$ in order to construct state
vectors which lead to tensor-factorization one has to invoke the \textit{split
property} which is known to break down in the case that the two regions touch
\cite{Haag}. The holographic projection compresses the vacuum fluctuations
into the lightlike direction and the relevant Hamitonian for the thermal
manifestation (including localization-entropy) is the lightlike dilation
generator which leaves $\partial W$ invariant.

The second result in the above list follows from this tensor-factorization;
transverse de-coupling implies additivity of extensive quantities. This is in
particular the case for the entropy and energy caused by the vacuum
fluctuations on the horizon; hence the contribution at the $x_{+}=0$ boundary
of $\partial W$ to the area density of the localization-entropy comes from
vacuum fluctuations in lightray direction. Since in the limit of sharp
localization we expect this density to become infinite, we introduce a
variable finite \textit{lightlike attenuation distance} $\varepsilon$ for
these vacuum polarization in order to study the limiting behavior for
$\varepsilon\rightarrow0.$

The computation of the leading $\varepsilon$-divergence is based on a theorem
\cite{S1} which states that a chiral system in a vacuum state $\Omega,$
localized on a halfline $\mathbb{R}_{+},$ can be unitarily mapped to the full
chiral algebra on $\mathbb{R}$ in a KMS thermal state $\Omega_{2\pi}$ at
temperature $2\pi$
\begin{equation}
(\mathcal{A}(\mathbb{R}_{+}),\Omega)\simeq(\mathcal{A}(\mathbb{R}%
),\Omega_{2\pi})\label{thm}%
\end{equation}
with the unitary equivalence being given in terms of a conformal map which
intertwines the dilation of the restricted system with the translation of the
unrestricted (two-sided) thermal system. This \textquotedblleft inverse Unruh
effect\textquotedblright\ for chiral theories (i.e. a heat bath system is
interpreted as a restricted vacuum system) is the key to the calculation of
localization-entropy. Using the intuition from statistical mechanics we would
conjecture that the entropy diverges proportional to the length i.e. as
$l\cdot s_{2\pi}$ where $s_{2\pi}$ is the density per length. The intertwining
map transforms the size $l$ into $\varepsilon=e^{-l}$ so that the area density
of the equivalent dilational system on the halfspace behaves as
\begin{equation}
s_{area}=\left\vert \ln\varepsilon\right\vert s_{2\pi}+finite,~~\varepsilon
\rightarrow0\text{ }%
\end{equation}
The correctness of this idea can be checked by approximating the linear system
by a sequence of finite systems using an \textquotedblleft
invariant\textquotedblright\ box approximation \cite{S1} in which the
divergent partition function of the translative system is approximated by a
rotational system in the limit of infinite temperature (interpreted as an
infinite radius whose size is related to $l$) associated with the high
temperature limit of the partition function associated to the Virasoro
generator $L_{0}.$ The calculation can be completed by the use of the chiral
temperature duality for the partition function of $\hat{L}_{0}=L_{0}-\frac
{c}{24}$ where the $c$ is the Virasoro constant. It is precisely this shift in
$L_{0}$ which gives the divergent $l$ factor and identifies the constant
$s_{2\pi}$ with $\frac{c}{6},$ so that the third result in the above list is
obtained. \ 

Chiral theories which originate from the chiral decomposition of conformal
two-dimensional theories come with an energy momentum tensor which contains
the Virasoro constant $c,$ but, as mentioned before, this is not necessarily
the case for Moebius covariant chiral theories which arise in holographic
lightfront projection. The fact that it is precisely the vacuum shift in
$\hat{L}_{0}$ which gives the expected $l$ divergence in the length
proportionality of the translative heat bath entropy may be taken as an
indication that the Virasoro structure continues to hold for chiral theories
from holographic projections. \ At the same time it gives an apparently new
thermal interpretation to the Virasoro constant $c$ in terms of properties of
a lightlike thermal system. It is this inverse Unruh effect in combination
with the simplification from the holographic projection which makes
localization-entropy a useful concept.

It is an interesting question to what extend such entropy considerations apply
to other null-surfaces including null-surfaces in Minkowski spacetime
different from $LF$. Since the thermal phenomenon under consideration is
caused by vacuum fluctuations near the boundary, one expects that the
localization-entropy has an area behavior with the same leading $\varepsilon
$-divergence as for the $LF$ null-surface. This is corroborated by the
structural theorem behind statement \textbf{1} (\ref{fac}) which secures the
absence of transverse vacuum polarizations on null surfaces. It begs the
further-going question whether there exist \textit{pointlike generators} which
obey commutations relations with transverse delta-functions similar to
(\ref{LF}) or (\ref{c.r.}). Whereas for conformally invariant bulk theories
one can rely on the global conformal equivalence of the wedge with a double
cone, we have not been able to prove that this holographic behavior extends to
massive theories on null-surfaces of double cones (the argument in \cite{S1}
in favor of such a result was incorrect). If there exist pointlike generators
on null-surfaces, they must lead to correlation functions containing the
quantum mechanical transverse delta-function factors because this is the only
way to guaranty the absence of transverse vacuum fluctuations in pointlike generators.

Null-surfaces as those associated with double cones are quite interesting
because one expects a quantum version of the classical Bondi-Metzner-Sachs
symmetry in such a case. Whereas the quantum symmetries are always defined
through their unitaries in the full Hilbert space (which only act
geometrically on the horizon) which is the same as the one for the bulk, the
BMS symmetry is only defined as an asymptotic transformation \cite{Wa}. As a
result of the more geometric nature (symmetry enhancement) of the action of
the modular group restricted to the horizon as compared to its fuzzy action on
massive bulk matter, one hopes to learn something about the modular action on
the bulk. obtain a solution of unsolved problem of an analytic understanding
of modular actions on localized bulk algebras. According to a recent proof of
the Bisognano-Wichmann theorem by \cite{Mund} based on a deep connection of
modular theory with the Haag-Ruelle scattering theory, one can also expect to
learn something about the action of the modular group on subwedge localized
bulk algebras from its action in free theories. These problems of extensions
of modular theory clearly go beyond the scope of an addendum and require a
separate publication.

In the case of conformal models one can try to compute generators for
double-cone holography by applying the relevant conformal transformation to
wedge generators. The conformal map from the $x_{0}-x_{3}$ wedge $W$ to the
radius=1 double cone $\mathcal{O}_{1}$ placed symmetrically around the origin
is%
\begin{align}
\mathcal{O}_{1} &  =\rho(W+\frac{1}{2}e_{3})-e_{3},\text{ }\rho(x)=-\frac
{x}{x^{2}}\\
W &  =\left\{  (x_{0},x_{\perp},x_{3})\ |~x_{3}>\left\vert x_{0}\right\vert
,x_{\perp}\in R^{2}\right\}  \nonumber
\end{align}
with $e_{3}$ being the unit vector in the 3-direction. Restricted to the
(upper) horizon $\partial W$ one obtains in terms of coordinates%
\begin{align}
\partial\mathcal{O}_{1} &  \ni(\tau,\vec{e}(1-\tau)),~\tau=\frac{t}%
{t+x_{\perp}^{2}+\frac{1}{4}},~\vec{e}=\frac{1}{x_{\perp}^{2}+\frac{1}{4}%
}(x_{\perp},\frac{1}{4}-x_{\perp}^{2})\\
&  where\text{ }\partial W=\left\{  (t,x_{\perp},t)\ |\ t>0,~x_{\perp}\in
R^{2}\right\}  \nonumber
\end{align}
If we use the unitary conformal transformation $\mathcal{A}(W)\rightarrow
\mathcal{A(O}_{1}\mathcal{)}$ not only on global generators for $\partial
\mathcal{A}(W)=\mathcal{A}(W)$ but also for their pointlike generationg fields
$A_{LF}$ (\ref{LF}), we obtain the desired compact transverse proportionality
factor $\sim\delta(\vec{e}-\vec{e}^{\prime})$ replacing $\delta(x_{\perp
}-x_{\perp}^{\prime})$ from the fact that the t-independent relation between
$\vec{e}$ and $x_{\perp}$ is that of a stereographic projection of $S^{2}$ to
$R^{2}.$ The presence of this factor corroborates the absence of vacuum
polarization in the above algebraic argument. The lightlike factor has the
expected qualitative behavior in terms of the variable $\tau$ and the
$W$-modular group $t\rightarrow e^{\lambda}t$ passes to the $\mathcal{O}_{1}$
modular automorphism
\begin{equation}
\tau\rightarrow\frac{-e^{-\lambda}(\tau+1)+1}{e^{-\lambda}(\tau+1)+1}%
\end{equation}
The transverse additive group passes via inverse stereographic transformation
to the transverse rotational group. Again it is important to realize that the
rotational symmetric two-point function of the pointlike generators of
$\partial\mathcal{O}_{1}$ is not identical to the bulk two-point function
restricted to $\partial\mathcal{O}_{1},$ in fact it is not even possible to
obtain generators for $\mathcal{A}(\partial\mathcal{O}_{1})$ by restricting
pointlike bulk generators so that the only possibility consists in conformally
transforming the lightfront generators (\ref{LF}).

The precise form of the lightray contribution to the local $\mathcal{O}_{1},$
including the proportionality factor for free fields, will be deferred to a
future paper in which also the important problem of the applicability of
ambient conformal maps restricted to null-surfaces of holographic projections
$\partial W\rightarrow\partial\mathcal{O}_{1}$ in case of massive bulk
theories will be investigated.

The divergence of localization-entropy in LQP is of course expected; since the
it is related to a chiral heat bath situation by a conformal transformation
which leads to a logarithmic distance parametrization which transforms the
chiral "volume" factor $L$ into the $\varepsilon$-attenuation distance factor
$\left\vert \ln\varepsilon\right\vert $ for the vacuum polarization cloud.
Hence the divergence is on the same structural-kinematical level as the large
volume divergence of heat bath entropy i.e. it is not related to any still
mysterious short distance divergencies caused by field generators in the bulk. 

The important difference to previous global entropy calculations based on
counting eigenstates of the standard time translation Hamitonian \cite{Bo-So}
is that in those calculations it is difficult to see the local origin of the
phenomenon which is responsible for the area proportionality. There are
general reasons which cast serious doubts on global calculations of vacuum
entropy and energy which count the contribution from the occupation of global
energy levels \cite{Ho-Wa}. Such calculations tend to violate the recently
formulated principle of local covariance of QFT in curved spacetime
\cite{B-F-V}. The present calculation does better on this issue since the
Hamiltonian is the same as the one responsible for the Hawking-Unruh effect
i.e. adapted to the invariance of the localization region and its horizon
(i.e. the Lorentz boost generator in case of $\mathcal{A}(W)$ for
generalizations to curved spacetime see next section). The theory is not
modified by momentum space cutoffs or in any other conceptually and
computationally uncontrollable way; one rather looks at a family of
localization entropies for the holographically projected matter with a fuzzy
boundary of a lightlike extension $\varepsilon$ at the edge of the horizon. In
particular this approach places into evidence that the rather universal
behavior of vacuum polarization clouds in an attenuation distance
$\varepsilon$ near boundaries with thermal consequences should not to be
blamed on the ultraviolet divergencies of particular pointlike fields (as they
show up in renormalization theory), but is rather a physical consequence of
the very peculiar operator-algebraic nature of causal localizability.
Localized operator algebras are known to be monads i.e. von Neumann factor
algebras of unique hyperfinite type III$_{1}$which have properties (e.g. no
normal pure states) which easily escape the quantum mechanical intuition
(which is limited to type I factor algebras \cite{Yng}). Some of the problems
black hole physics allegedly has with the foundations of QT may find their
explanation in these peculiarities.

\section{Extension to black hole horizons}

It has been known for some time that QFT in CSTv with a bifurcated Killing
horizon (bKh) lead to a similar situation as wedges and their horizons
\cite{Kay-Wa}. A presentation in the setting of operator algebras in curved
space time for such a situation which turns out to be also suitable for the
calculation of localization-entropy has been given in \cite{G-L-R-V} following
prior work in \cite{Sew}\cite{Su-Ve}. The main difference to the Minkowski
situation is that at the start one only has a Killing symmetry and some
assumptions about its geometrical action which guaranty the existence of a
bKh. As a substitute for the vacuum state one needs a state vector $\Omega$
which is invariant under the action of the Killing symmetry; for the
Schwarzschild-Kruskal black-hole spacetime this would be the Hartle-Hawking
state \cite{Wald}. In the analogy to the Minkowski spacetime the Killing
symmetry corresponds to the wedge-preserving L-boost whose projection onto the
bKh is the candidate for a dilation. It comes then as quite a surprise that it
is possible to embed this Killing symmetry as a dilation into a full
3-parametric Moebius group, including a positive energy lightlike translation.
In contradistinction to the $LF$ holography these objects are only defined on
the bKh; to be more precise they are implemented by operators in the Hilbert
space of the QFT which leave $\Omega$ invariant; but they act geometrically
only in their restriction to bKh. This is an illustration of the gain of
symmetries through holographic projection. The definition of the net of
algebras on the horizon is similar to the Minkowski case in that the wedge
regions $W$ possess a bKh analog; instead of the use of the Wigner little
group to resolve the transverse locality structure on $\partial W_{a,b}$ one
intersects $\mathcal{A}(\partial W_{a,b})$ with bulk algebras whose
localization region contains the subregion of $\partial W_{ab}$ of given
transverse extension (tubular neighborhoods) to which one wants to localize
the algebra. The only subtle part of the construction of the holographic
projection to the horizon is the construction of the positive energy
translation from the theory of modular inclusions. For this and other related
details we refer to \cite{G-L-R-V} which is the most authoritative account of
this matter.

Having a horizon with a chiral lightray structure, the transverse
factorization along the edge of the bKh and the area proportionality of
entropy follows as in the lightfront case. There is no change in the divergent
part of the area density; the so-called \textit{surface gravity} which enters
the relation between the dilation and the translation modifies only the next
to leading (constant) term in the $\varepsilon$-expansion. Equating this
entropy density with Bekenstein's classical area density would of course
relate the size of the vacuum polarization cloud to the Planck length. The
present purely structural approach cannot resolve the question whether
different c-values belonging to a different quantum matter content really
occur. But in case that the holographically projected matter does lead to
different $c$-values this would lead to a problematization with the matter
independent pure gravity-based Bekenstein density which is usually interpreted
as including the contribution coming from quantum matter. In view of the fact
that the state of black hole radiation of a collapsing star is not really an
equilibrium state \cite{Haw}\cite{Fr-Ha}, one is of course entitled to doubt
the relevance of localization-entropy or classical Bekenstein entropy and
speculate that the recent operator-algebraic concept based on an entropy flux
\cite{Flow} is more appropriate than the static entropy concept of equilibrium
statistical mechanics.

As already remarked at the end of the previous section, the difference to the
existing entropy calculations is conceptually significant. Whereas previous
field theoretic calculations incorporate the finiteness of the Bekenstein law
by introducing a global momentum space cutoff (which amounts to an
uncontrollable modification of the theory), the present approach interprets
the area behavior as a manifestation of absence of transverse vacuum
fluctuations, leaving the numerical matching of the area densities to a better
future understanding (or the key to QG) of how the back-reaction of quantum
gravity determines the size $\varepsilon$ of the halo of vacuum fluctuations
at the edge of the horizon i.e. it is more in the spirit of a quasiclassical
approximation which maintains the original QFT setting. Most of the
astrophysical and cosmological vacuum effects (e.g. the cosmological constant)
of recent interests have been calculated by quantum mechanical level
occupation arguments i.e. by a procedure which does not comply with the
principle of local covariance \cite{Ho-Wa}. We believe that the present
approach, based on holographic projection to null-surfaces, is exempt from
this criticism.

Finally it is important to stress that the notion of holography which is in
widespread use in the literature (originally introduced by 't Hooft) is
different from the one in the present work. The main difference that the
former is thought of as leading to a complete holographic image from which the
full bulk theory can be recovered without any additional information about the
bulk. Such an invertible \textquotedblleft duality\textquotedblright\ relation
between bulk and its holographic image, if possible at all, is expected by
some people to arise \textquotedblleft magically\textquotedblright\ from the
still elusive quantum gravity. The present notion of holography onto
null-surfaces applies to QFT in curved spacetime and lacks this uniqueness of
inversion (unless one adds a few additional properties which refer to the
ambient spacetime and enforce uniqueness). Even in the case of wedge
localization it is not possible to reconstruct the localization structure
within a wedge $W$ from that of its horizon $\partial W.$ But adding the
knowledge about actions of Poincar\'{e} transformations outside the
7-parametric subgroup $\mathcal{G}$ of $LF$ allows the reconstruction of the
full net of wedge algebras and via intersection the full net of finitely
localized bulk algebras and their possible generating pointlike fields. Since
even in case of free fields there is no local connection between bulk fields
and holographic field generators; the intervention of operator algebras is an
essential aspect of holography. Apart from the structural results used in the
present work, the operator algebra theory (in particular the Tomita-Takesaki
modular theory) as one needs it the physical setting is still very much in its
infancy; there is presently no good intuitive understanding of how
geometric-physical properties in spacetime are related to the posititioning of
hyperfinite type III$_{1}$ factor algebras in a common Hilbert space.

The relation of the conjectured quantum gravitational holography on
null-surfaces to the present one is reminiscend to that of the conjectured
Maldacena \cite{Ma} gravity-gauge correspondence to Rehren's algebraic AdS-CFT
correspondence \cite{Reh}. The full invertibility (which justifies the use of
the word correspondence) of the latter is the proven result of the exceptional
fact that the causal shadows cast by the regions of a \textquotedblleft
conformal brane\textquotedblright\ boundary at infinity of AdS can be
intersected and subsequently transported anywhere (as a result of the maximal
shared symmetry of bulk and boundary) to form the net on AdS; in the Maldacena
setting AdS-CFT correspondence is interpreted as a conjectured result of the
kind of quantum gravity inherent in string theory.

\textit{Acknowledgements}: I thank my colleagues Joao Barata and Walter
Wreszinski for their hospitality during a visit of the USP where part of this
work was done. I also acknowledge the financial support of the FAPESP during a
one year stay at the Mathematical Physics Group of the Physics Department of
the USP.

\end{document}